**Similarity Solutions of the Einstein-Maxwell Equations with One Killing Vector**

Elliot Fischer[1]
 6 Leigh Drive, Madison, N.J. 07940

**Abstract**
New similarity variables are introduced for the Einstein - Maxwell equations with one Killing vector that reduce the non-linear partial differential equations in three independent variables to ordinary differential equations. These similarity variables are the extensions of those previously found for the Einstein-Maxwell equations with two Killing vectors [1,2,3]. The resulting equations are then solved, providing new solutions of the Einstein - Maxwell equations with one Killing Vector.

**1.     Introduction**
The problem of finding exact solutions to the Einstein or Einstein-Maxwell equations has always been very difficult.  Solutions for space-times with only one Killing vector field (or three independent variables) are particularly difficult to come by, due to the combination of the nonlinear nature of the equations and the appearance of more than two independent variables.  In this paper we present solutions that use similarity techniques to reduce the nonlinear partial differential equations to nonlinear ordinary differential equations, which may then be solved.  The technique can be viewed as an extension of similar techniques previously developed for the Einstein equations with two Killing vectors [2-5].  The applicability of this technique implies that the Einstein-Maxwell equations with one Killing vector possess a symmetry group similar to that with two Killing Vectors that can be exploited to find solutions.

**2.     Einstein- Maxwell Field Equations for Spacetimes with One Killing Vector and Assumed Metric**
In this section, we review the Einstein equations for spacetimes admitting one Killing vector.  We use the notation from [4] . The assumed form of the metric is:

$$-ds^2 = \varepsilon[\, e^{2U} (dx^k + a f_\alpha\, dx^\alpha)^2 + a^2 e^{-2U} \gamma_{\alpha\beta} dx^\alpha dx^\beta\,] \tag{1}$$

where   $\varepsilon = +/- 1 = \operatorname{sgn}(g_{kk})$; $\qquad(2)$

a is an aribtrary scale factor; k is some particular value of 0,1 , 2 , 3; Greek letters take on all values of 0,1,2,3 except k; and all metric coefficients are independent of $x^k$. This metric thus admits a Killing vector $\xi^l = \delta^l_k$.

We now define $\gamma^{\alpha\beta}$ as the inverse of the 3-dimensional metric $\gamma_{\alpha\beta}$,

$$\gamma^{\alpha\beta} \gamma_{\alpha\beta} = \delta^\beta_\delta , \tag{3}$$

and $\Sigma^\alpha{}_{\beta\gamma}$ and $P_{\alpha\beta}$ as the Christoffel symbols and Ricci tensor, respectively, obtained from the $\gamma_{\alpha\beta}$. Harrison also defines differential operators of first and second order:

$$\Delta_1 (F) = \gamma^{\alpha\beta} F_{,\alpha} F_{,\beta} , \tag{4}$$
$$\Delta_1 (F,G) = \gamma^{\alpha\beta} F_{,\alpha} G_{,\beta} , \tag{5}$$
$$\Delta_2 (F) = \gamma^{\alpha\beta} (F_{,\alpha\beta} - \Sigma^\gamma{}_{\alpha\beta} F_{,\beta} ), \tag{6}$$

where the comma denotes ordinary partial differentiation ,and F and G are any functions of the coordinates. The semicolon always represents covariant differentiation with respect to the $\gamma_{\alpha\beta}$. Harrison continues the calculations for the Einstein- Maxwell equations in [4]. We summarize his results, using our assumptions on the metric, as follows.

Assume the $\gamma_{\alpha\beta}$ are constant, corresponding to a conformally flat background metric. In addition, Harrison introduces 2 potentials for the electromagnetic field, A and B. We assume B to be 0. With these assumptions, the Einstein-Maxwell equations become

$$\Delta_2 (A) = 2 \Delta_1 (U,A) \tag{7}$$

$$\Delta_2 (U) = - \varepsilon e^{-2U} \Delta_1 (A) \tag{8}$$

$$P_{\alpha\beta} = 2 U,_\alpha U,_\beta + 2 \varepsilon e^{-2U} A,_\alpha A,_\beta \tag{9}$$

where $P_{\alpha\beta}$ is the Ricci tensor of the background metric $\gamma_{\alpha\beta}$. Since we have chosen the background metric to be conformally flat, the Ricci tensor is zero and equation (9) becomes

$$0 = 2 U,_\alpha U,_\beta + 2 \varepsilon e^{-2U} A,_\alpha A,_\beta \tag{10}$$

We now must solve (7), (8), and (10). Looking at (10) and choosing $\varepsilon = -1$, (10) is satisfied if

$$A = e^U \tag{11}$$

Upon substitution into (7) and (8), they both reduce to the equation

$$\Delta_2 (U) = \Delta_1 (U) \tag{12}$$

## 3. Solutions and Similarity Variables

We now assume that the Killing vector is in the direction of the angle φ in cylindrical coordinates. This makes the operators $\Delta_2 (U)$ and $\Delta_1 (U)$ independent of φ and (12) can be written as

$$U,_{\rho\rho} + U,_{\rho}/\rho + U,_{zz} - U,_{tt} = (U,_{\rho})^2 + (U,_{z})^2 - (U,_{t})^2 \tag{13}$$

where ρ and z are the usual cylindrical coordinates, and t is time.

Eq. (13) has a form similar to the equations with 2 Killing Vectors discussed in [1-5], with respect to both the left hand side and the non-linear nature of the right hand side.

### 3.1 Similarity Variable

Following Fischer [1,2], where the similarity variable σ = ρ/t reduced similar Einstein-Maxwell equations with 2 Killing vectors to ordinary differential equations, we now suggest that the similarity variable σ = ρ/(z+t) will reduce (13) to an ordinary differential equation. With this substitution, (13) becomes, after some simplification,

$$U'' + U'/\sigma = (U')^2 \tag{14}$$

Thus, by use of the similarity variable, the partial differential equation in 3 independent variables (13) has been reduced to an ordinary differential equation in 1 independent variable. (14) may be further solved to yield

$$U = -\ln(C_1 \ln\sigma + C_2) \tag{15}$$

where $C_1$ and $C_2$ are integration constants. The electromagnetic potential A may then be found from (11) by exponentiating U. This solution is apparently new and does not appear in [6], the definitive treatise on exact solutions of the Einstein and Einstein-Maxwell Equations.

### 3.2 Generalized Similarity Solution

Again following [1,2,3] we extend the generalized similarity variable σ = $\rho^2 + z^2$

to $\sigma = \rho^2 + z^2 - t^2$. Substitution of this form into (13) yields the ordinary differential equation

$$U'' + 2U'/\sigma = (U')^2 \tag{16}$$

Again, by use of the similarity variable, the partial differential equation in 3 independent variables (13) has been reduced to an ordinary differential equation in 1 independent variable. (16) may be further solved to yield

$$U = -\ln(C_1 \sigma^{-1} - C_2) \tag{17}$$

where $C_1$ and $C_2$ are constants of integration. Again, this solution is apparently new and does not appear in [6], the definitive treatise on exact solutions of the Einstein and Einstein-Maxwell Equations.

## 4. Conclusions

We have shown that the Einstein-Maxwell field equations with one Killing vector and a conformally flat background metric can be reduced to ordinary differential equations using extensions of the similarity and generalized similarity variables for the Einstein and Einstein-Maxwell equations with two Killing Vectors.

This implies that the Einstein - Maxwell equations with one Killing vector have an isovector or generalized isovector that lead to these solutions. Based on this, it is reasonable to postulate that the Einstein - Maxwell equations with one Killing vector have more similarity variables, since they are similar in structure to the Einstein - Maxwell equations with 2 Killing vectors, which are known to have a large isovector group [5].

Future research will try to relax the assumption of a conformally flat background metric and determine if the same or similar similarity variables can reduce the resulting equations.

**Acknowledgements**
The author would like to thank B. Kent Harrison for many correspondences and the suggestion of using a conformally flat background metric, which led to many simplifications. Thanks also go to Frank Estabrook for his constant encouragement. The author would also like to thank Mr. Frank DeLisi's sixth grade class at Central Avenue School, Madison. N.J. in 1994 for helping to suggest this problem.